\documentclass[nopreprintline,12pt,authoryear]{elsarticle}

\usepackage{lineno}

\usepackage{amsmath,amsfonts}
\usepackage{algorithmic}
\usepackage{algorithm}
\usepackage{array}
\usepackage{textcomp}
\usepackage{stfloats}
\usepackage{url}
\usepackage{verbatim}
\usepackage{graphicx}
\usepackage{cite}
\usepackage{subfigure}
\usepackage{enumerate}
\usepackage{hhline}
\usepackage{bm}
\usepackage{comment}
\usepackage{upgreek}
\usepackage{xcolor}
\usepackage{booktabs}
\usepackage{lineno}

\newcommand{\given}{\,|\,}
\newcommand{\T}{\top}

\newcommand{\calD}{\mathcal{D}}
\newcommand{\bC}{\mathbf{C}}

\newcommand{\by}{\mathbf{y}}
\newcommand{\bX}{\mathbf{X}}
\newcommand{\bx}{\mathbf{x}}

\newcommand{\bV}{\mathbf{V}}

\newcommand{\calU}{{\cal U}}
\newcommand{\bH}{\mathbf{H}}

\newcommand{\bM}{\mathbf{M}}

\newcommand{\bI}{\mathbf{I}}
\newcommand{\bD}{\mathbf{D}}

\newcommand{\mb}{\mathbf{m}}
\newcommand{\bmu}{\boldsymbol{\mu}}

\newcommand{\beps}{\boldsymbol{\epsilon}}
\newcommand{\bbeta}{\boldsymbol{\beta}}

\newcommand{\bomega}{\boldsymbol{\omega}}

\newcommand{\btheta}{\boldsymbol{\theta}}

\graphicspath{{./arxiv/}}

\begin{document}

\begin{frontmatter}



\title{Bayesian Modeling of Incompatible Spatial Data: A Case Study Involving Post-Adrian Storm Forest Damage Assessment}


\author[inst1]{Lu Zhang}
\author[inst2]{Andrew O. Finley}
\author[inst3]{Arne Nothdurft}
\author[inst5]{Sudipto Banerjee}

\affiliation[inst1]{organization={Division of Biostatistics, Department of Population and Public Health Sciences, Keck School of Medicine},
            addressline={University of Southern California}, 
            city={Los Angeles},
            postcode={90033}, 
            state={CA},
            country={USA}}

\affiliation[inst2]{organization={Departments of Forestry and Probability \& Statistics},
            addressline={Michigan State University}, 
            city={East Lansing},
            postcode={48824}, 
            state={Michigan},
            country={USA}}

\affiliation[inst3]{organization={Department of Forest and Soil Sciences},
            addressline={University of Natural Resources and Life Sciences}, 
            city={Vienna},
            postcode={1190}, 
            state={Vienna},
            country={Austria}}

\affiliation[inst4]{organization={Department of Biostatistics,
Fielding School of Public Health},
            addressline={University of California}, 
            city={Los Angeles},
            postcode={0095}, 
            state={California},
            country={USA}}

\begin{abstract}
Modeling incompatible spatial data, i.e., data with different spatial resolutions, is a pervasive challenge in remote sensing data analysis. Typical approaches to addressing this challenge aggregate information to a common coarse resolution, i.e., compatible resolutions, prior to modeling. Such pre-processing aggregation simplifies analysis, but potentially causes information loss and hence compromised inference and predictive performance. To avoid losing potential information provided by finer spatial resolution data and improve predictive performance, we propose a new Bayesian method that constructs a latent spatial process model at the finest spatial resolution. This model is tailored to settings where the outcome variable is measured on a coarser spatial resolution than predictor variables---a configuration seen increasingly when high spatial resolution remotely sensed predictors are used in analysis. A key contribution of this work is an efficient algorithm that enables full Bayesian inference using finer resolution data while optimizing computational and storage costs. The proposed method is applied to a forest damage assessment for the 2018 Adrian storm in Carinthia, Austria, that uses high-resolution laser imaging detection and ranging (LiDAR) measurements and relatively coarse resolution forest inventory measurements. Extensive simulation studies demonstrate the proposed approach substantially improves inference for small prediction units.
\end{abstract}


\end{frontmatter}




\section{Introduction}

Synthesis of data collected at different spatial resolutions is a component of most contemporary analyses that aim to learn about environmental processes or patterns using georeferenced ground measurements and information collected by possibly different remotely stationed sensors. This synthesis often takes the form of a regression-like model in which the environmental variable of interest is defined as some function of the remotely sensed variables. Here, we refer to the environmental variable as the outcome and remotely sensed variables as predictors.

Data with different spatial resolutions are often referred to as \emph{incompatible} spatial data, and methods for coupling such data within statistical models has received substantial attention within the geospatial, ecological, and geostatistical literature, see, e.g., field-specific foundational papers by \citet{Atkinson2000scale}, \citet{Dungan2002scale}, and \citet{gotway2002combining}. These and related works recognize that valid inference about the process that generates outcome observations requires the outcome and predictors share the same spatial resolution or model accommodations are made to appropriately relate across disparate resolutions. 

In the remote sensing literature, creating compatible spatial data, i.e., data of the same spatial resolution, involves downscaling or upscaling. Downscaling refers to an increase in spatial resolution, e.g., decrease in pixel size of remotely sensed data. Conversely, upscaling refers to a decrease or coarsening in spatial resolution, e.g., increasing pixel size via aggregation. For valid inference, downscaling and upscaling methods should accommodate uncertainty quantification. 

In the geostatistical literature, matching disparate resolutions within a statistically valid framework is called the change-of-support (COS) problem and is often addressed using kriging-based methods \citep{gotway2002combining, kyriakidis2004geostatistical}. Such kriging methods involve constructing a spatial model for an underlying process at a selected fine spatial resolution and treating the observation on a coarser region as the aggregation of the fine process in that region. In the remote sensing and geospatial literature, statistical downscaling refers to a subset of COS methods \citep{atkinson2013downscaling}.

In many settings, outcome observations are measured at a spatial resolution finer than that of remotely sensed predictors. This is mostly due to the often prohibitively high cost of collecting in situ outcome measurements and relatively coarse resolution of commonly used sensors or gridded data products, see applied examples in \citet{Peng2017}, \citet{Zheng2021}, \citet{Berrocal2012}, \citet{Berrocal2020}, and \citet{Zhang2021}. In such settings, COS methods focus on downscaling predictor measurements to match outcome measurements.

Increased access to high spatial resolution remotely sensed data, e.g., laser imaging detection and ranging (LiDAR) and hyperspectral, means that predictors can often be measured at a spatial resolution finer than that of the outcome measurements. This setting differs from the COS problems noted above because the outcome and predictor spatial resolutions are reversed and, hence, require different considerations when formulating appropriate models. Such settings are now common in many forestry applications, where forest outcomes measured on fixed-area inventory plots are related to LiDAR predictors available at a scale finer than that of a given inventory plot. For example, a recent study by \citet{nothdurft2021estimating} aimed to model a forest outcome measured on relatively large inventory plots and predictors derived from high-resolution LiDAR data---there were approximately 140 LiDAR predictor measurements for each outcome measurement. To address their incompatible spatial data issue, \citet{nothdurft2021estimating} used a pre-processing upscaling step to aggregate LiDAR measurements to match the inventory plot area, then fit their proposed regression models and made subsequent predictions. Although their approach provided rapid model parameter estimation, subsequent prediction accuracy and precision might not be optimal, especially when a prediction unit's area is considerably smaller than the inventory plot. Their approach to the COS problem was non-statistical in nature and hence there might be room for improved prediction and associated uncertainty quantification.

In this study, we propose a new Bayesian model for settings where the outcome variable is measured on a coarser resolution than the predictor variables. We pursue a Bayesian inferential framework because it adds flexibility to the model's specification, parameter estimation, and prediction \citep{gelfand2001change, wikle2005combining, banerjee2014hierarchical, cressie2015statistics, bradley2016bayesian}. We motivate the method development using a reanalysis of the \citet{nothdurft2021estimating} study. Our approach to this COS problem constructs a spatial model at the finest spatial resolution (e.g., that of the LiDAR predictor data). Additionally, we develop an efficient implementation algorithm for the proposed model. We demonstrate through simulation studies that our approach leads to improved prediction performance in some settings. The remainder of this article is organized as follows. Section~\ref{sec:data} provides an overview of the \citet{nothdurft2021estimating} study and data. The proposed COS modeling approach is outlined in Section~\ref{sec:model}, followed by the implementation algorithm in Section~\ref{sec:implementation}. We conduct a comparative analysis of the new algorithm and the algorithm presented by \citet{nothdurft2021estimating} through simulation studies in Section~\ref{sec:simulation}. Section~\ref{sec:blowdownAnalysis} presents the results of our data analysis using the newly proposed algorithm. We conclude the article with a discussion and ideas for future work in Section~\ref{sec:discussion}.

\subsection{Study area and blowdown data}\label{sec:data}

\begin{figure}[!ht]
\begin{center}
 \includegraphics[height=.5\textwidth,trim={0cm 0cm 0cm 0cm},clip]{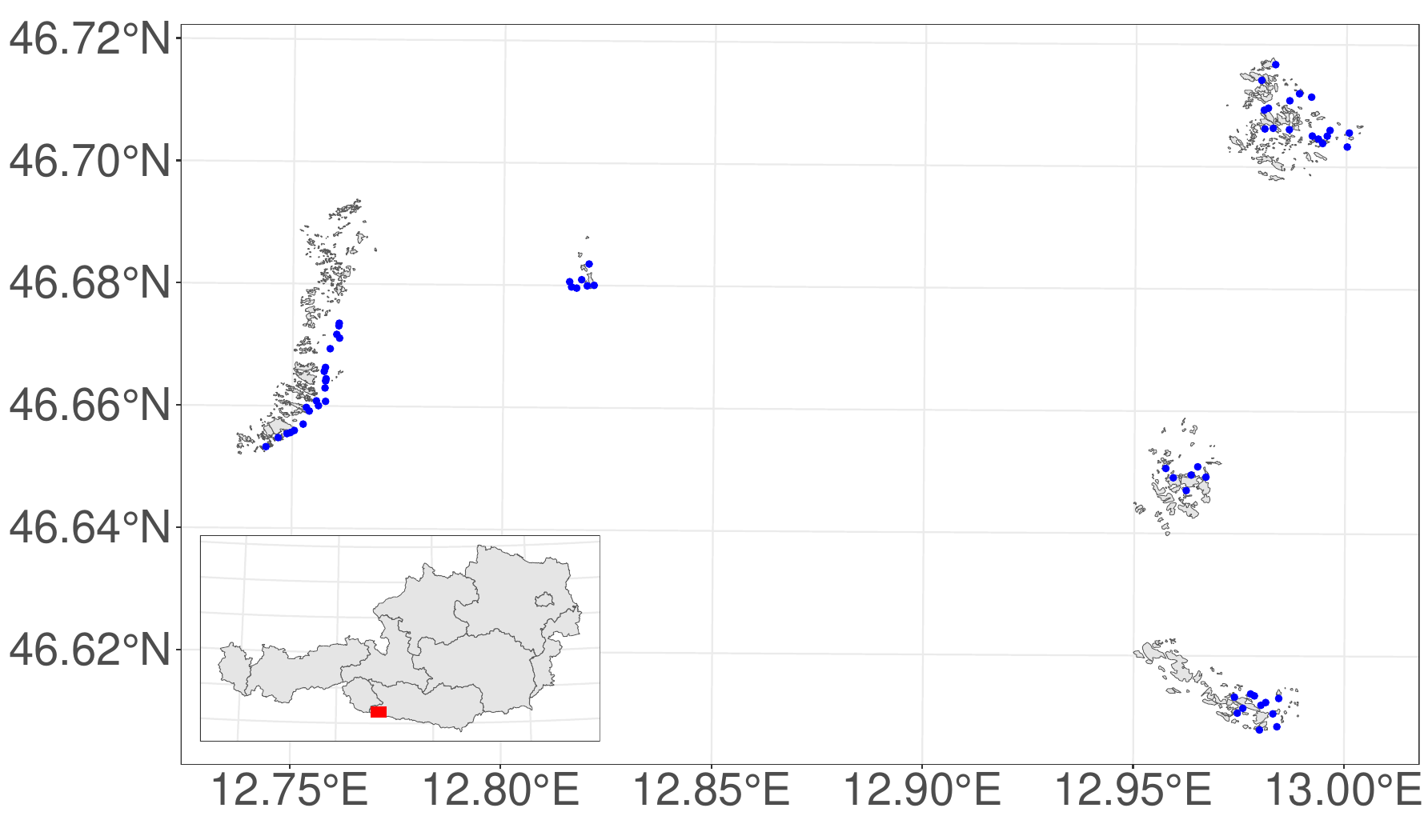}
 \caption{Map inset is the location of the study area in Southern Carinthia, Austria (red point). Storm damage areas (polygons) and inventory plots (blue points) in the study area.} \label{fig:StudySite1}
\end{center}
\end{figure}

The \citet{nothdurft2021estimating} study focused on an area in the upper Gail Valley in the southern Austrian federal state of Carinthia that is close to the Dellach Forest Research and Training Center of the Institute of Forest Growth and the Institute of Forest Engineering of the University of Natural Resources and Life Sciences Vienna (Figure~\ref{fig:StudySite1} inset). On October 28, 2018, storm Adrian brought wind gust speeds of 130\,km/h across Carinthia and produced heavy rainfalls of 627\,\,l/m$^2$ within 72 hours at the Pl{\"o}ckenpass meteorological station, located near the study area \citep{Zimmermann2018}. 

The storm damages caused by hurricane Adrian mostly occurred in the Hermagor administrative district (Bezirk). In total, 564 blowdown areas were identified and delineated using high-resolution aerial images provided by the Carinthian Forest Service. The damages were concentrated in five distinct sub-regions: Frohn, Laas, Liesing, Mauthen, and Ploecken (Figure~\ref{fig:StudySite1}). The total area of the damaged forests was 212.3\,ha. Summary statistics for the blowdown areas are provided in Table~\ref{tab:SumStatAreaPlot}.

The study's inferential goal was to predict what 2020 growing stock timber volume would have been---if not destroyed by Adrian in late October 2018---in each blowdown and summarized over blowdowns at the sub-region and regional scales. 

\begin{table*}[!ht]
\caption{Summary statistics of the digitized blowdowns and inventory plot data.\label{tab:SumStatAreaPlot}}
\centering
{\tiny
\begin{tabular}{lrrrrrrrrrrrr}
& \multicolumn{6}{c}{Blowdown areas} & \multicolumn{6}{c}{Inventory plots}\\
\cmidrule(lr){2-7} \cmidrule(lr){8-13}
& & \multicolumn{5}{c}{Area (ha)} & & \multicolumn{5}{c}{Growing stock volume (m$^3$/ha)}\\
\cmidrule(lr){3-7} \cmidrule(lr){9-13}
Site & count & mean & med & sd & min & max & $n_b$ & mean & med & sd & min & max \\ 
\midrule
Frohn & 273 & 0.230 & 0.065 & 0.600 & 0.004 & 7.231 & 21 & 593 & 579 & 227 & 188 & 979 \\ 
Laas & 152 & 0.362 & 0.138 & 0.565 & 0.004 & 2.727 & 17 & 785 & 772 & 333 & 142 & 1382 \\ 
Liesing & 5 & 0.581 & 0.348 & 0.621 & 0.115 & 1.607 & 7 & 689 & 698 & 67 & 563 & 757 \\ 
Mauthen & 62 & 0.621 & 0.190 & 1.075 & 0.013 & 4.969 & 6 & 762 & 748 & 323 & 403 & 1220 \\ 
Ploecken & 72 & 0.738 & 0.242 & 1.724 & 0.008 & 12.322 & 11 & 773 & 773 & 136 & 528 & 968 \\ 
\midrule
Total & 564 & 0.376 & 0.110 & 0.892 & 0.004 & 12.322 & 62 & 705 & 730 & 256 & 142 & 1382 \\  
\end{tabular}
}
\end{table*}

\begin{figure}[!ht]
\begin{center}
	\includegraphics[width=1\textwidth,trim={0cm 3.5cm 0cm 3.5cm},clip]{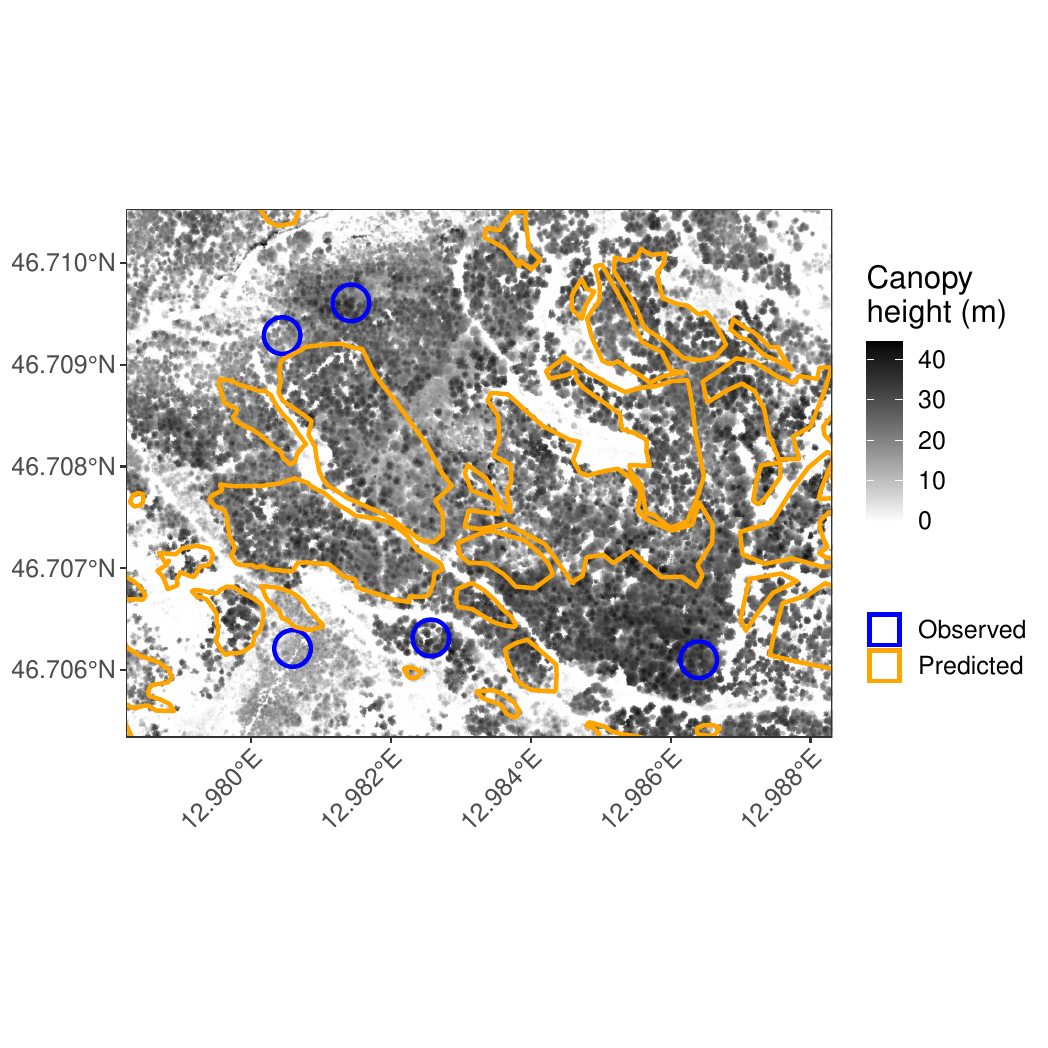}
	\caption{A portion of the Laas sub-region depicting location and extent of observed circular inventory plots (blue) where timber growing stock volume was measured, blowdowns where growing stock volume prediction is desired (orange), and 3 m $\times$ 3 m spatial resolution forest canopy height data used as a predictor for growing stock volume.} \label{plotAndPredictionUnitsLass}
\end{center}
\end{figure}

Because there were no pre-hurricane forest inventory data available for the study area, an ad hoc field measurement campaign was conducted in May 2020 to collect growing stock timber volume samples that proved suitable for modeling the volume loss in the blowdown areas. As detailed in \citet{nothdurft2021estimating}, $n_b$=62 inventory plots were sampled in undamaged forest adjacent to the blowdowns and having forest structure similar to that of the blowdowns prior to the hurricane. Each plot was a 20 m radius circle (0.126 ha) within which tree measurements were used to produce the outcome measurements of growing stock timber volume (m$^3$/ha). Figure~\ref{fig:StudySite1} shows the location of all $n_b$ plots, and Figure~\ref{plotAndPredictionUnitsLass} shows a close-up view of the plots (blue circles) and their proximity to blowdowns (orange polygons) for the Laas sub-region. Summary statistics for the inventory plot data are given in Table~\ref{tab:SumStatAreaPlot}.

Lidar predictors were derived from gridded canopy height data provided by Carinthian Forest Service in a 3\,m $\times$ 3\,m pixel resolution originating from airborne laser scanning data collected in 2012. While many canopy height metrics can be computed using the LiDAR point cloud within a given pixel, following \citet{nothdurft2021estimating} we consider only mean canopy height as the LiDAR predictor in the subsequent reanalysis. There are 139.63 forest canopy height pixel measurements within each inventory plot, see, e.g., Figure~\ref{plotAndPredictionUnitsLass}. It is the spatial resolution incompatibility between growing stock volume plot measurements and forest canopy height pixel measurements we address using our proposed COS approach detailed in the next section.  

\section{Proposed COS methods}\label{sec:model}

Here we developed the predictive spatial model to analyze the incompatible spatial data in this study. As stated in Section~\ref{sec:data}, we use the 2012 LiDAR measurements summarized on a 3\,m $\times$ 3\,m resolution grid as the predictors, and the outcome variable is the growing stock volume measured on 20\,m radius plots collected in 2020. As illustrated in Figure~\ref{plotAndPredictionUnitsLass}, blowdowns have varied sizes and irregular shapes. On the 3\,m $\times$ 3\,m resolution grid, there are 8,661 and 235,909 pixels that overlap with observed plots and prediction blowdowns, respectively. However, the number of observed plots and prediction blowdowns are only 62 and 564, respectively. Each observed plot covers 140 3\,m $\times$ 3\,m pixels, and the smallest prediction blowdown covers 5 3\,m $\times$ 3\,m pixels.

Assume the predictors are recorded on a fine grid $\{A_{i}: i = 1, \ldots, n_a \}$. We use $\bx(A_i) = (x_1(A_i), \ldots, x_p(A_i))^\top$ to denote the $p$-dimensional vector of the predictor variables observed on unit $A_{i}$, and $\{A_{i}: i = 1, \ldots, n_a \}$ is the finest scale upon which we plan to build our models. In our study, we fit a regression coefficient for each sub-region and $\bx(A_i)$ is a vector of length $p = 5$ comprising the LiDAR measured mean canopy height, i.e., the $j$-th element in $\bx(A_i)$ equals the LiDAR point cloud distribution's mean over $A_i$ when $A_i$ is located in sub-region $j$, and it equals zero otherwise. The finest scale in our study is the 3\,m $\times$ 3\,m grid. The outcome variable is observed on plots ($\{B_l: l = 1, \ldots, n_b\}$). We use $y(B_l)$ to denote the measured outcome over $B_l \subset \calD$, where $\calD \subset \mathbb{R}^d$ is the study domain. In our study,  $y(B_l)$ is growing stock volume (m$^3$/ha) on the $l$-th plot and $n_b$=62.
On the finest scale we assume the spatial model
\begin{linenomath*}
\begin{equation}\label{eq: fine_model}
		y(A_i) = \beta_0 + \omega(A_i) + \bx(A_i)^\top \bbeta_1 + \epsilon(A_i)\;,
\end{equation}
\end{linenomath*}
where $\beta_0$ is the intercept, $\omega(A_i)$ is the latent spatial process, $\bbeta_1$ is the vector of regression coefficients associated with predictors $\bx(A_i)$, and $\epsilon(A_i)$ is a white noise process. Note that $\epsilon(A)$ cannot legitimately arise from the integral of a point-referenced white noise process $\epsilon(s)$, i.e., as $\int_A \epsilon(s)ds$ since this integral is not well-defined. Hence, $\epsilon(A_i)$'s are modeled independently as white noise over the finite collection of regions as its support. We model $\omega(A_i)$ through a zero-centered Gaussian process with an exponential correlation function (although, of course, other valid spatial correlation functions could be used). Denote $\bomega_A = (\omega(A_1), \ldots, \omega(A_{n_a}))^\top$, then $\bomega_A$ follows a multivariate normal distribution 
$\bomega_A \sim \mbox{MVN}(0, \sigma^2 \bC_\phi)$,
where the $(i,j)$-th element of the correlation matrix $\bC_\phi$ is defined through the exponential correlation function $\exp(-\phi \cdot d(s_i, s_j))$ with $s_i, s_j$ the centroids of $A_i$ and $A_j$, $d(s_i, s_j)$ the Euclidean distance between $s_i, s_j$, and the parameter $\phi$ controlling the rate of correlation decay. Since all pixels on the grid have the same area, we further assume $\epsilon(A_i)  \stackrel{ind}{\sim} \mbox{normal}(0, \tau^2)$,
where $\tau^2$ is the variance of the white noise on the finest scale.
Akin to block kriging \citet{Wackernagel2003}, the predictive average for each target unit $B_l, l = 1, \ldots, n_b$ is a weighted average of multiple point predictions on the fine grid within the target unit extent. 
Under the assumption that the underlying model at the finest resolution follows \eqref{eq: fine_model}, for each $l$ we have,
\begin{linenomath*}
\begin{equation}\label{eq: aggre_y}
\begin{split}
		& y(B_l) = \frac{1}{|B_l|} \sum_{i = 1}^{n_a} y(A_i)\int_{B_l \cap A_i} ds \approx \sum_{i = 1}^{n_a} \frac{|B_l \cap A_i|}{|B_l|} y(A_i) \\
		&\quad = \beta_0 + \sum_{i = 1}^{n_a} h_{li} \omega(A_i)+ \sum_{i = 1}^{n_a} h_{li}  \bx(A_i)^\top \bbeta_1  + \sum_{i = 1}^{n_a} h_{li} \epsilon(A_i)\;,
\end{split}
\end{equation}
\end{linenomath*}
where $h_{li} = \frac{|B_l \cap A_i|}{|B_l|}$ and $|B|$ represents the area of $B \subset \calD$. Denote $\by_B = (y(B_1), \ldots, y(B_{n_b}))^\top$. Let $\bH_{BA}$ be an $n_b \times n_a$ matrix whose $(i,j)$-th element is $h_{ij}$, we have
\begin{linenomath*}
\begin{equation}\label{eq: matrix_data_model}
\begin{aligned}
    \by_B = \bH_{BA}\bomega_A + \bH_{BA}
	\bX \bbeta + \bH_{BA}\beps_A 
\end{aligned}
\end{equation}
\end{linenomath*}
where $\bX = \begin{bmatrix}1 &\cdots& 1\\
\bx(A_1)& \cdots &\bx(A_{n_a}) \end{bmatrix}^\top$, $\bbeta = \begin{bmatrix} \beta_0 \\ \bbeta_1 \end{bmatrix} $  $\beps_A = (\epsilon(A_1), \ldots, \epsilon(A_n))^\top$.
We assign priors
\begin{linenomath*}
\begin{equation}\label{eq: priors}
\begin{aligned}
\bbeta  &\sim \mbox{Normal}(\bmu_\beta, \bV_{\beta} )\;, \; \tau^2 \sim \mbox{Inverse-Gamma}(t_{\tau}, s_{\tau})\; , \\
\sigma^2 &\sim \mbox{Inverse-Gamma}(t_{\sigma}, s_{\sigma})\;,\; \phi \sim \mbox{Uniform}(a_\phi, b_\phi)\;,
\end{aligned}
\end{equation}
\end{linenomath*}
with prefixed parameters $\{\bmu_\beta, \bV_{\beta}, t_{\tau}, s_{\tau}, t_{\sigma}, s_{\sigma}, a_\phi, b_\phi\}$. 
The shape and scale of the inverse gamma prior for $\tau^2$ ($\sigma^2$) are specified by $t_{\tau}$ ($t_{\sigma}$) and $s_{\tau}$ ($s_{\sigma}$), respectively, while $a_\phi, b_\phi$ denote the support of the uniform distribution. Combining the priors \eqref{eq: priors}, the data model \eqref{eq: matrix_data_model}, and model for $\omega_A$, we build up the Bayesian hierarchical model for our study. 
For generalization, any proper priors for $\tau, \sigma, \phi$ works, and the prior of $\bbeta$ can be replaced by a flat prior. 

Predictions are achieved using \eqref{eq: fine_model}~and~\eqref{eq: aggre_y}. Assume $B^u \subset \calD$ is the unobserved prediction unit. We pick pixels from the fine grid $\{A_{i}: i = 1, \ldots, n_a \}$ that cover $B^u$ and denote the picked pixels as $\{A^u_i: i = 1, \ldots, n_u\}$. Assume we have the posterior samples of $\bomega_A$ on the pixels that cover the observed units $\{B_j: j = 1, \ldots, n_b\}$. 
Through \eqref{eq: fine_model} we see that $\omega(A^u_i)$ and $\by_B$ are independent conditional on $\bomega_A$ and the joint posterior distribution $p( \omega(A^u_i),  \bomega_A, \bbeta, \sigma^2, \tau^2, \phi \given \by_B)$ equals
\begin{linenomath*}
\begin{equation}\label{eq: pred_A}
   	 p(\omega(A^u_i) \given \bomega_A, \bbeta, \sigma^2, \tau^2, \phi) \times p(\bomega_A, \bbeta, \sigma^2, \tau^2, \phi \given \by_B)\;. 
\end{equation}
\end{linenomath*}
Since the latent spatial process $\omega(A_i)$ is modeled through a Gaussian process, $\omega(A^u_i) \given \bomega_A, \sigma^2, \phi$ follows a Gaussian distribution with a closed form. After obtaining the posterior samples of all parameters $\{\bomega_A, \bbeta, \sigma^2, \tau^2, \phi\}$ in the Bayesian hierarchical model, we can generate posterior samples of $\omega(A^u_i)$ using the conditional distribution of $\omega(A^u_i) \given \bomega_A, \sigma^2, \phi$. Finally, by replacing $B_l$ by $B^u$ in \eqref{eq: aggre_y}, we can generate posterior samples $y(B^u)$ based on posterior samples of $\omega(A^u_i)$, $\bbeta$, and $\tau^2$. 

\subsection{Implementation method}\label{sec:implementation}

The model and corresponding prediction algorithm based on \eqref{eq: pred_A} involves posterior sampling of the high dimensional parameter $\bomega_A$. However, sampling on a parameter space including $\bomega_A$ through Markov chain Monte Carlo (MCMC) can be challenging. In our study, $n_a$ (the length of $\bomega_A$) is 8,661. The computational burden and storage requirement at each iteration of MCMC when including $\bomega_A$ in the parameter space are in the order of $\mathcal{O}(n_a^3)$ and $\mathcal{O}(n_a^2)$, respectively, which is a heavy cost. Besides, the high-dimensional posterior distribution on the joint space may not be well explored by MCMC sampling, resulting in slow convergence rate and poor mixing. For the sake of efficiency, we develop an algorithm to avoid sampling the high dimensional parameters $\bomega_A$. To be specific, we first conduct MCMC sampling on the low-dimensional space $\{\sigma^2, \tau^2, \phi \}$. Then we recover the posterior samples of $\bomega_B = \bH_{BA}\bomega_A$ and $\bbeta$. In the last stage, we use the posterior samples of $\btheta = \{\sigma^2, \tau^2, \phi, \bomega_B, \bbeta \}$ to obtain the predictive inference. 

Following the model design in the preceding subsection, the posterior density $p(\bbeta, \bomega_B, \sigma^2, \tau^2, \phi \given \by_B)$ is proportional to
\begin{linenomath*}
\begin{multline}\label{eq: log-density}
p(\tau^2, \sigma^2, \phi)\times \mbox{MVN}(\bomega_B \given 0, \sigma^2 \bC_B(\phi)) \times \mbox{MVN}(\bbeta \given \bmu_\beta, \bV_\beta) \\
   \times \mbox{MVN}(\by_B \given \bomega_B + \bH_{BA} \bX \bbeta, \tau^2 \bD_h)\;,
\end{multline}
\end{linenomath*}
where $\bD_h = \bH_{B,A}\bH_{B,A}^{\T}$ is a diagonal matrix (this follows from the structure of the pixels within the blocks) with $l$-th diagonal element equal to $\sum_{i = 1}^{n_a} h^{2}_{li}$, and the $(i,j)$-th element in $\bC_B(\phi)$ equals 
\begin{linenomath*}
\begin{equation}\label{eq: CB_element}
    \sum_{A_l \cap B_i \neq \emptyset, A_k \cap B_j \neq \emptyset} h_{il} h_{kj} \exp(- \phi \cdot d(s_l, s_k))\;.
\end{equation}
\end{linenomath*}
Therefore, the marginal posterior on the low-dimensional space $\{\sigma^2, \tau^2, \phi\}$ follows
\begin{linenomath*}
\begin{equation}\label{eq: marginal_posterior}
\begin{aligned}
p(\sigma^2, \tau^2, \phi \given \by_B ) 
    &\propto \mbox{MVN}(\by_B \given \bH_{BA} \bX \bmu_\beta, \sigma^2 \bC_B(\phi) +\\
    \bH_{BA} \bX &\bV_\beta \bX^\top \bH_{BA}^\top + \tau^2 \bD_h) \times p(\tau^2, \sigma^2, \phi)
\end{aligned}
\end{equation}
\end{linenomath*}
Furthermore, with a flat prior for $\bbeta$, i.e. $p(\bbeta) \propto 1$, 
\begin{linenomath*}
\begin{align}\label{eq: marginal_post_flat_b}
&p(\sigma^2, \tau^2, \phi \given \by_B ) \propto p(\tau^2, \sigma^2, \phi) \times \nonumber\\
&\quad \frac{|\bV_{\beta}^\ast|^{1/2}}{|\bV^\ast|^{1/2}} \exp[-\frac{1}{2} \{\by_B^\top \bV^{\ast-1} \by_B - \bmu_\beta^{\ast\top}\bV_{\beta}^{\ast-1}\bmu_\beta^{\ast} \}] \;,
\end{align}
\end{linenomath*}
where 
\begin{linenomath*}
\begin{align*}
\bV_\beta^{\ast-1} &= \bX^\top \bH_{BA}^\top \bV^{\ast-1}\bH_{BA}\bX,\\ 
\bV^{\ast} &= \sigma^2 \bC_B(\phi) + \tau^2 \bD_h, \\
\bmu_\beta^\ast &= \bV_\beta^{\ast} \bX^\top \bH_{BA}^\top \bV^{\ast-1}\by_B,\text{ and}\\ 
\bmu_\beta^{\ast\top}\bV_{\beta}^{\ast-1}\bmu_\beta^{\ast} &= \by_B^\top \bV^{\ast-1} \bH_{BA}\bX\bV_\beta^{\ast}\bX^\top \bH_{BA}^\top \bV^{\ast-1}\by_B. 
\end{align*}
\end{linenomath*}
The derivation of \eqref{eq: marginal_posterior} and \eqref{eq: marginal_post_flat_b} is provided in the Appendix. Given that $n_b$, the length of $\bomega_B$, is 62 in the study, the computational cost of $p(\sigma^2, \tau^2, \phi \given \by_B)$ is dominated by the generation of $\bC_B(\phi)$ whose cost is in the order of $\mathcal{O}(n_a^2)$. To further reduce the computational burden, we use a tapering kernel \citep{wendland1995piecewise, gneiting2002compactly} to taper the spatial correlation function $\exp(-\phi \cdot d(s_l, s_k))$ in \eqref{eq: CB_element} to zero when $d(s_l, s_k)$ is beyond a certain range. This method of introducing sparsity is known as covariance tapering, see, e.g.,  \citet{furrer2006covariance}, \citet{kaufman2008covariance}, and \citet{du2009fixed}. A popular choice of tapering kernel is $C_\gamma(s_l, s_k) = (1 - d(s_l, s_k)/\gamma)^4_{+}(1 + 4d(s_l, s_k)/\gamma)$ where $\gamma$ is the distance beyond which the correlation becomes zero. The exponential correlation function noted in the preceding development is replaced by the tapered correlation function 
\begin{linenomath*}
\begin{equation*}
C_{tap}(s_l, s_k ; \phi) = \exp(-\phi \cdot d(s_l, s_k)) \cdot C_\gamma(s_l, s_k)\;.
\end{equation*}
\end{linenomath*}
The distance $\gamma$ is a prefixed parameter. In our study, we set $\gamma$ at 1.8 km to reduce the spatial correlation of the plots from different sub-regions into zero while maintaining as many non-zero correlations among the observations within the same sub-region as possible. We implement the marginal posterior model in probabilistic programming language like Stan \citep{Stan2017} to generate posterior samples. In our implementation, we generate 4 MCMC chains, each with 500 iterations for warm-up and 500 iterations for sampling, and then select $G = 200$ posterior samples ($\{\sigma^{2(g)}, \tau^{2(g)}, \phi^{(g)}\}, g = 1, \ldots, G$) by keeping one of every $10$ iterations of the MCMC chains. 

In the algorithm's second stage, we recover the posterior samples of $\{\bbeta, \bomega_B\}$ using the posterior samples of $\{\sigma^2, \tau^2, \phi\}$. Since we have 
\begin{linenomath*}
\begin{equation*}
p(\bbeta, \bomega_B, \sigma^2, \tau^2, \phi \given \by_B ) = p(\bbeta, \bomega_B \given \sigma^2, \tau^2, \phi, \by_B ) p(\sigma^2, \tau^2, \phi \given \by_B)\;,
\end{equation*}
\end{linenomath*}
and from \eqref{eq: log-density} the conditional posterior distribution has a closed form (see proofs in the Appendix)
\begin{linenomath*}
\begin{equation}\label{eq: cond_posterior_B}
\bbeta, \bomega_B \given \sigma^2, \tau^2, \phi, \by_B \sim \mbox{MVN}( \bM\mb, \bM)\;,
\end{equation}
\end{linenomath*}
where 
\begin{linenomath*}
{\small \begin{align*}
    \bM^{-1} &=\begin{bmatrix}
    \bX^\top \bH_{BA}^\top \\ \bI_B \end{bmatrix} (\tau^2 \bD_h)^{-1} [\bH_{BA}\bX, \bI_B] + \begin{bmatrix} \bV_\beta^{-1} & 0 \\ 0 & \frac{1}{\sigma^2}\bC_B^{-1}(\phi) \end{bmatrix} \;, \;\\ 
    \mb &=
    \begin{bmatrix} \bX^\top \bH_{BA}^\top\\ \bI_B \end{bmatrix}(\tau^2 \bD_h)^{-1}\by_B + \begin{bmatrix}  \bV_\beta^{-1}\bmu_\beta\\ 0 \end{bmatrix}\;.
\end{align*}}
\end{linenomath*}
When assigning flat prior for $\bbeta$, we just need to replace $\bV_\beta^{-1}$ in the above equation by a zero matrix.
This step involved Cholesky decomposition of the matrix $\bM^{-1}$. 
We can compute the Cholesky decomposition with posterior sample $\{\sigma^{2(g)}, \tau^{2(g)}, \phi^{(g)}\}$ and generate posterior sample $\bbeta^{(g)}, \bomega_B^{(g)}$ for each $g$. 

Lastly, we derive a closed form of the conditional posterior predictive distribution and use it to obtain the posterior predictive inference. Assume $\calU = \{B^u_1, \ldots, B^u_{n_p}\}$ contains the regions for prediction. For $B^u \in \calU$, since we just want to estimate $y(B^u)$, we obtain posterior prediction based on the posterior samples of $\omega(B^u) = \sum_{i = 1}^{n_u} h_{i}^u \bomega(A_i^u)$ where $h_{i}^u =\frac{|B^u \cap A_i^u|}{|B^u|}$. Let $\bomega^u_B = (\omega(B^u_1), \ldots, \omega(B^u_{n_p}))^\top$,
\begin{linenomath*}
\begin{align}\label{eq: wu_pred}
\bomega^u_B \given \bomega_B, \phi \sim \mbox{MVN}(& \bC_{BU}^\top \bC_B^{-1}(\phi) \bomega_B,\nonumber\\ 
&C_{UU} - \bC_{BU}^\top \bC_B^{-1}(\phi)\bC_{BU})\;,
\end{align}
\end{linenomath*}
where the $(l, k)$-th element of $\bC_{BU}$ 
equals 
\begin{linenomath*}
\begin{equation*}
\sum_{\{s_i: s_i \cap B_l \neq \emptyset\}, \{s_j^u: s_j^u \cap B_k^u \neq \emptyset\}} h_{li} h_{kj}^u C_{tap}(s_i, s_j^u; \phi),
\end{equation*}
\end{linenomath*}
the $(l, k)$-th element of $\bC_{UU}$ equals
\begin{linenomath*}
\begin{equation*}
\sum_{\{s_i^u: s_i^u \cap B_l^u \neq \emptyset\}, \{s_j^u: s_j^u \cap B_k^u \neq \emptyset\}} h_{li}^u h_{kj}^u C_{tap}(s_i^u, s_j^u; \phi),
\end{equation*}
\end{linenomath*}
and $h_{kj}^u =  \frac{|B_k^u \cap A_j^u|}{|B_k^u|}$.
Then the posterior prediction $y(B^u)$ for $B^u \in \calU$ simply follows 
\begin{linenomath*}
\begin{equation}\label{eq: post_pred}
\mbox{N}(\beta_0 + \bomega(B^u) + \sum_{i = 1}^{n_u} h_{i}^u \bx(A_i^u)^\top \bbeta_1, \tau^2 \sum_{i = 1}^{n_u} h^{u2}_{i})\;.
\end{equation}
\end{linenomath*}
With the posterior samples $\theta^{(g)} = \{\sigma^{2(g)}, \tau^{2(g)}, \phi^{(g)}, \bbeta^{(g)}, \bomega_B^{(g)}\}, g = 1, \ldots G$ obtained from the first two stages, we can generate posterior predictive sample $\bomega^{u(g)}_B$ and $\{y(B^u)^{(g)}: B^u \in \calU \}$ for each $\theta^{(g)}$ by \eqref{eq: wu_pred} and \eqref{eq: post_pred}. We provide a summary of the sampling algorithm in Algorithm~\ref{alg: implement}.

\begin{algorithm}
\caption{Sampling for the proposed Bayesian spatial model}\label{alg: implement}
\begin{algorithmic}[1]
\STATE Generate posterior samples $\{\sigma^{2(g)}, \tau^{2(g)}, \phi^{(g)}\}_{g = 1}^G$ through Stan using target density defined by \eqref{eq: marginal_posterior} or \eqref{eq: marginal_post_flat_b}
\FOR{$g = 1$ to $G$}
\STATE Sample $\bbeta^{(g)}, \bomega_B^{(g)}$ with  $p(\bbeta, \bomega_B \given \sigma^{2(g)}, \tau^{2(g)}, \phi^{(g)})$ defined by \eqref{eq: cond_posterior_B}
\STATE Sample $\bomega_B^{u(g)}$ with $p(\bomega_B^{u} \given \bomega_B^{(g)}, \phi^{(g)})$ defined by \eqref{eq: wu_pred}
\STATE Sample $y(B^u)^{(g)}$ from \eqref{eq: post_pred} with $\bbeta = \bbeta^{(g)}$, $\tau^2 = \tau^{2(g)}$, and $\omega(B^u) = \omega(B^u)^{(g)}$
\ENDFOR
\end{algorithmic}
\end{algorithm}

\section{Simulation study}\label{sec:simulation}
We anticipate the proposed COS approach will yield improved predictive inference particularly when prediction units (e.g., blowdowns) have irregular shapes and when they have an area comparable to or smaller than the observation units (e.g., inventory plots). We illustrate this feature through simulation studies in this section. For comparison, we include the prediction approach, referred to as ``Block'', used in \citet{nothdurft2021estimating} as a competitor. The Block approach partitions each blowdown into pixels that have the same area as the fixed-area inventory plots (i.e., 0.126 ha). Each pixel is indexed using its centroid, and predictor variables are computed as an average of the LiDAR values over its extent. The Block approach uses the spatial model described in \eqref{eq: fine_model}, but at the coarsest area resolution defined by the plot area, i.e., replacing $A$s with $B$s. The predictive average over plots for prediction for both algorithms are computed through \eqref{eq: aggre_y}, a weighted average of the point predictions on the pixels overlaps with the plot for prediction. Essentially, the major difference between the Block approach and our proposed approach is that our approach builds models on a much finer grid and it remains scalable for massive data using the proposed implementation algorithm. We refer to our proposed approach as ``COS'', because it models observed outcome and predictors with incompatible supports.

\begin{figure}[!ht]
\begin{center}
	\subfigure[]{\includegraphics[width=0.45\textwidth,trim={0cm 0.5cm 0.5cm 1.8cm},clip]{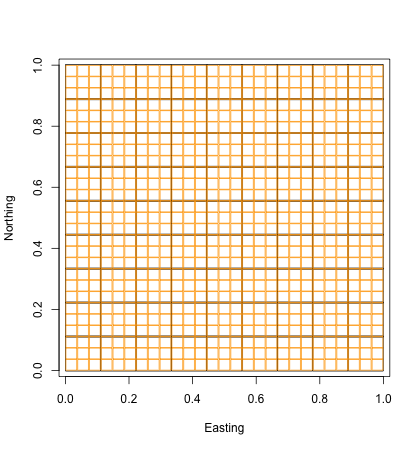}\label{fig:simgrid}}
	\subfigure[]{\includegraphics[width=0.45\textwidth,trim={0cm 0.5cm 0.5cm 1.8cm},clip]{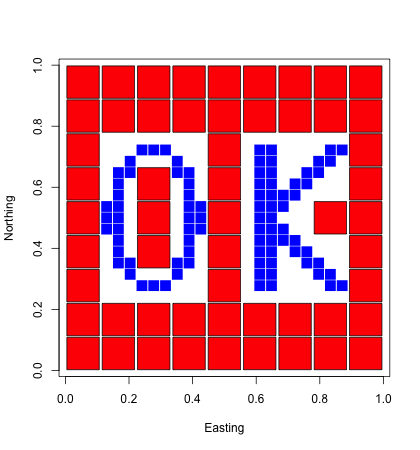}\label{fig:simOK}}
	\caption{(a) Coarse (black) and fine (orange) grids used in the simulation study. (b) Illustration of observed units (red squares) and predict units (the ``O'' and ``K'' as indicated by the blue squares) in the simulation study.} \label{fig:sim}
\end{center}
\end{figure}

In this simulation, we generated data through the spatial regression model \eqref{eq: fine_model} on a 1/27 $\times$ 1/27 resolution grid on a unit square $[0, 1]^2$. The predictor $\bx(A_i)$ for each pixel is a single value sampled from a standard normal. We pick $\beta_0 = 1$, $\beta_1 = 5$, $\tau^2 = 1.0$, and the latent spatial process $\omega(A_i)$ follows a zero-centered Gaussian process with exponential correlation function with $\sigma^2 = 2.0$ and $\phi = 5.0$. 
We provide an illustration of the grid in Figure~\ref{fig:simgrid}, where the grid in orange and black indicate the fine and coarse grids, respectively. As shown in Figure~\ref{fig:simgrid}, each observation on the coarse grid averages 9 observations on the fine grid.  In Figure \ref{fig:simOK} we use the red squares to indicate the observed pixels in the coarse grid.  The ``O'' and ``K'' in blue are two prediction units.  Unit ``O'' covers 38 in the fine grids. Unit ``K'' covers 48 pixels in the fine grids.

For our proposed COS approach, the model is fit on the fine grid, and we set $\gamma = 0.6$ since the correlation of the latent process $\omega(A_i)$ on two pixels drop below $0.05$ when the distance between the centroids of the two pixels is greater than $0.6$. For the Block approach, the model is fit on the coarse grid, and compute the predictor as the average of the predictor values on the fine grid over each pixel on the coarse grid. We do not do any approximation like covariance tapering in the Block approach. We assigned the same priors for both models. The prior for $\bbeta$ was set to be a zero-centered Gaussian distribution with covariance matrix equal to $1000\cdot \bI_2$. And the remaining hyperparameters in the priors \eqref{eq: priors} were $t_\tau = t_\sigma = 2$ $s_\tau = s_\sigma = 2$, $a_\phi = 0.006$ and $b_\phi = 30$. We generated 200 posterior predictive samples of the average outcome on ``O'' and ``K'' through both approaches. The simulation was repeated 100 times. 

\begin{table*}[!ht]
\caption{Summary of root mean squared prediction error (RMSPE), median absolute prediction error (MPE), and 95\% credible interval coverage and width for the ``OK'' simulation study. Values are the average over 100 replicates.}\label{tab:simsum}
\centering
\resizebox{\textwidth}{!}{
\begin{tabular}{lcccccccc}
\toprule
  & RMSPE O & RMSPE K & MPE O & MPE K & CI cover O & CI cover K & CI width O & CI width K\\
\midrule
Block & 0.69 & 0.69 & 0.40 & 0.46 & 0.69 & 0.74 & 1.45 & 1.68\\
COS & 0.37 & 0.47 & 0.25 & 0.36 & 0.91 & 0.96 & 1.34 & 2.06\\
\bottomrule
\end{tabular}}
\end{table*}

The predictive performance of the COS and Block approaches was compared using the root mean squared prediction error (RMSPE), the median absolute prediction error (MPE) and the empirical coverage rates of the 95\% credible interval (CI cover). For each replicate, we calculated the prediction error for each prediction unit, ``O'' and ``K'', as the difference between the posterior mean and the true value. The RMSPE and MPE for each prediction unit were then computed based on the 100 prediction errors across all replicates. Table~\ref{tab:simsum} summarizes these metrics for each prediction unit, as well as the average width of the credible interval for each unit (CI width). The COS approach results in lower RMSPEs (0.37 for ``O'' and 0.47 for ``K'') compared to the Block approach (0.69 for ``O'' and 0.69 for ``K'').  The median of the absolute prediction errors generated by the Block approach for ``O'' (``K'') were 1.57 (1.30) times that of the COS approach. 
Among the 100 simulations, the COS approach yielded average coverage rates of 91\% and 96\% for ``O'' and ``K'', respectively. The corresponding coverage rates from the Block approach drop to 69\% and 74\%. It is noteworthy that prediction unit CI width for ``O'' generally decreases when transitioning from the Block to COS approach. While this trend is reversed for prediction unit ``K''. These trends suggest the COS approach effectively uses the fine resolution predictor information to provide a richer and more nuanced understanding of uncertainty. Although the Block approach was over 20 times ($\sim$3s vs. $\sim$65s) faster than the COS approach in this simulation, we conclude the COS approach provides better uncertainty quantification than the Block approach according to Table~\ref{tab:simsum}.

The preceding simulation study demonstrates that the COS approach offers more reliable predictions for smaller areas. A natural question to ask is how the COS approach performs when the prediction unit area varies. To investigate this, we extended the initial simulation study by expanding the predictive units to include the white pixels surrounding plots ``O'' and ``K'' in Figure~\ref{fig:simOK}. The updated summary table is presented in Table~\ref{tab:simsum2}. Comparing Table~\ref{tab:simsum}~and~\ref{tab:simsum2}, we observe that the performance of the Block approach and COS approach converges as the resolution of the prediction units becomes larger. Based on these simulation studies, we anticipate the performance of the Block and COS approaches will further converge when the prediction unit area increases. Considering the speed benefit, the Block approach is preferred for predictions over large areas, while the COS approach is more beneficial for predictions in smaller areas.

\begin{table*}[!ht]
\caption{Summary of root mean squared prediction error (RMSPE), median absolute prediction error (MPE), and 95\% credible interval coverage and width for the large area ``OK'' simulation study.  Values are the average over 100 replicates.}\label{tab:simsum2}
\centering
\resizebox{\textwidth}{!}{
\begin{tabular}{lcccccccc}
\toprule
  & RMSPE O & RMSPE K & MPE O & MPE K & CI cover O & CI cover K & CI width O & CI width K\\
\midrule
Block & 0.30 & 0.43 & 0.22 & 0.27 & 0.97 & 0.88 & 1.30 & 1.54\\
COS & 0.28 & 0.37 & 0.20 & 0.24 & 0.93 & 0.95 & 1.06 & 1.60\\
\bottomrule
\end{tabular}}
\end{table*}

\section{Blowdown comparative analysis}\label{sec:blowdownAnalysis}

We fit the COS and Block models to the blowdown data described in Section~\ref{sec:data} with the inferential goal of predicting growing stock timber volume loss for each blowdown, sub-region, and region. Prior specifications and posterior inference for parameters and blowdown predictions followed methods in Section~\ref{alg: implement} and Block modeling presented \citet{nothdurft2021estimating}. To validate the algorithms, we conducted a 10-fold cross-validation test across the 62 observed plots and summarized the results in Table~\ref{tab:CV}. The two candidate approaches were assessed based on 10-fold cross-validation prediction RMSPE, MPE, and continuous rank probability score (CRPS) \citep{gneiting2007strictly}. CRPS is a preferable measure of predictive skill because it favors models with both high accuracy and precision. Lower RMSPE and CRPS indicate better predictive performance. Additionally, we computed the percentage of holdout observations covered by their corresponding 95\% credible interval. Model with empirical coverage percentages close to the nominal 95\% level is favored. As shown in Table~\ref{tab:CV}, the Block and COS approaches provide comparable predictive performance in the cross-validation test. Because the sizes of the observed plots are the same as those of the prediction units (that is, the left-hand plots), these results are consistent with the findings of our simulation study. Next, we consider the COS approach to enhance predictions for each blowdown, particularly for those with smaller areas.

\begin{table*}[!ht]
\caption{10-fold cross-validation prediction diagnostics. The diagnostics were calculated using predictions from a 10-fold cross-validation on the observed data. The CI Cover column indicates the percentage of 95\% posterior predictive credible intervals that cover the observed held-out values. The MPE column shows the median absolute prediction error.}\label{tab:CV}
\centering
\resizebox{0.6\textwidth}{!}{
\begin{tabular}{lrrrr}
\toprule
  & CI cover & RMSPE & MPE & CRPS\\
\midrule
Block & 0.97 & 157.818 & 118.166 & 5506.217\\
COS & 0.92 & 161.911 & 101.995 & 5596.635\\
\bottomrule
\end{tabular}}
\end{table*}

\begin{figure}[!ht]
	\centering
		\includegraphics[width=0.8\textwidth,trim={0cm 0cm 0cm 0cm},clip]{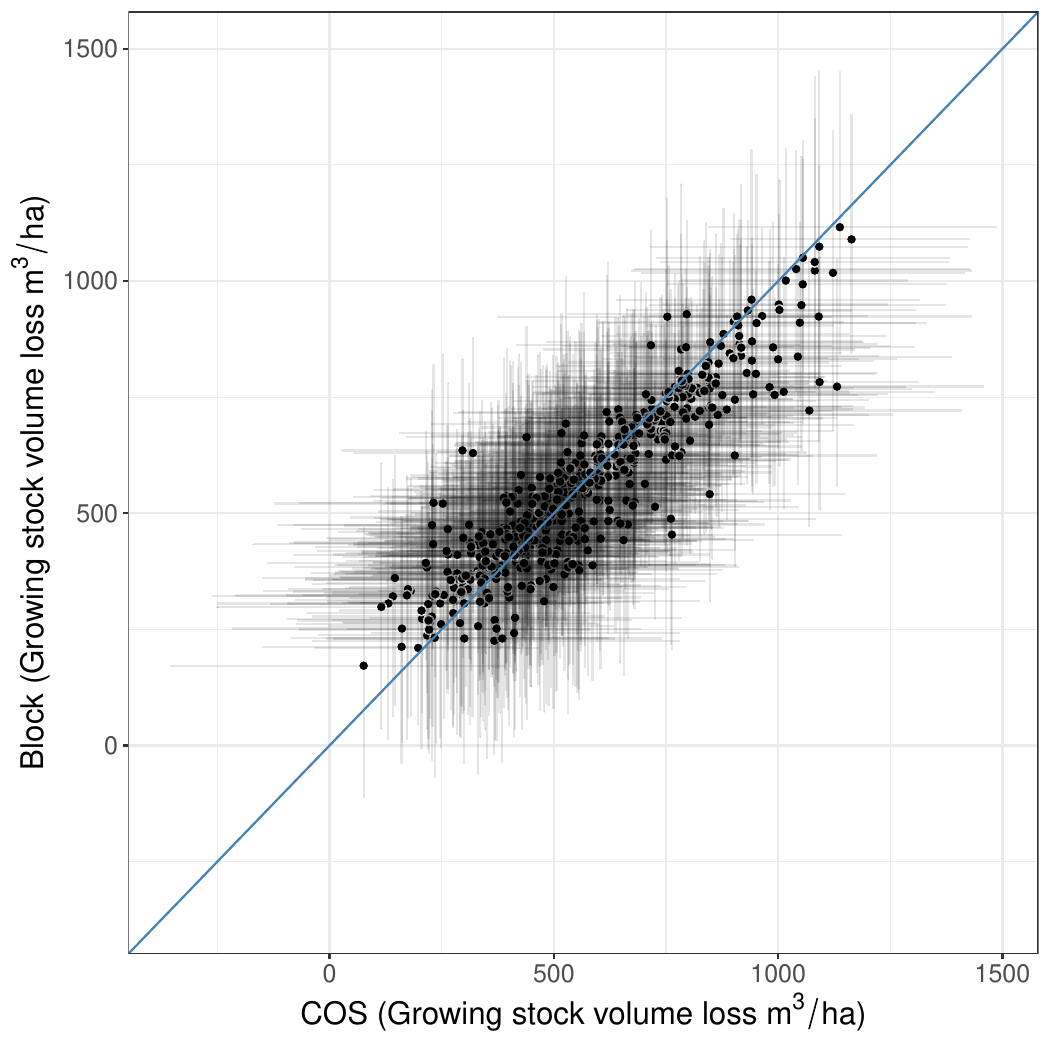}
		
	\caption{COS vs. Block growing stock volume loss posterior predictive distribution (PPD) summaries for each of the 546 blowdowns. Points are PPD medians and horizontal and vertical lines are associated 95\% credible intervals for COS and Block, respectively.} \label{fig:realCOSvsBlock}
\end{figure}

\begin{figure}[ht!]
	\centering
	\includegraphics[width=0.8\textwidth,trim={0.1cm 0cm 0.0cm 0cm},clip]{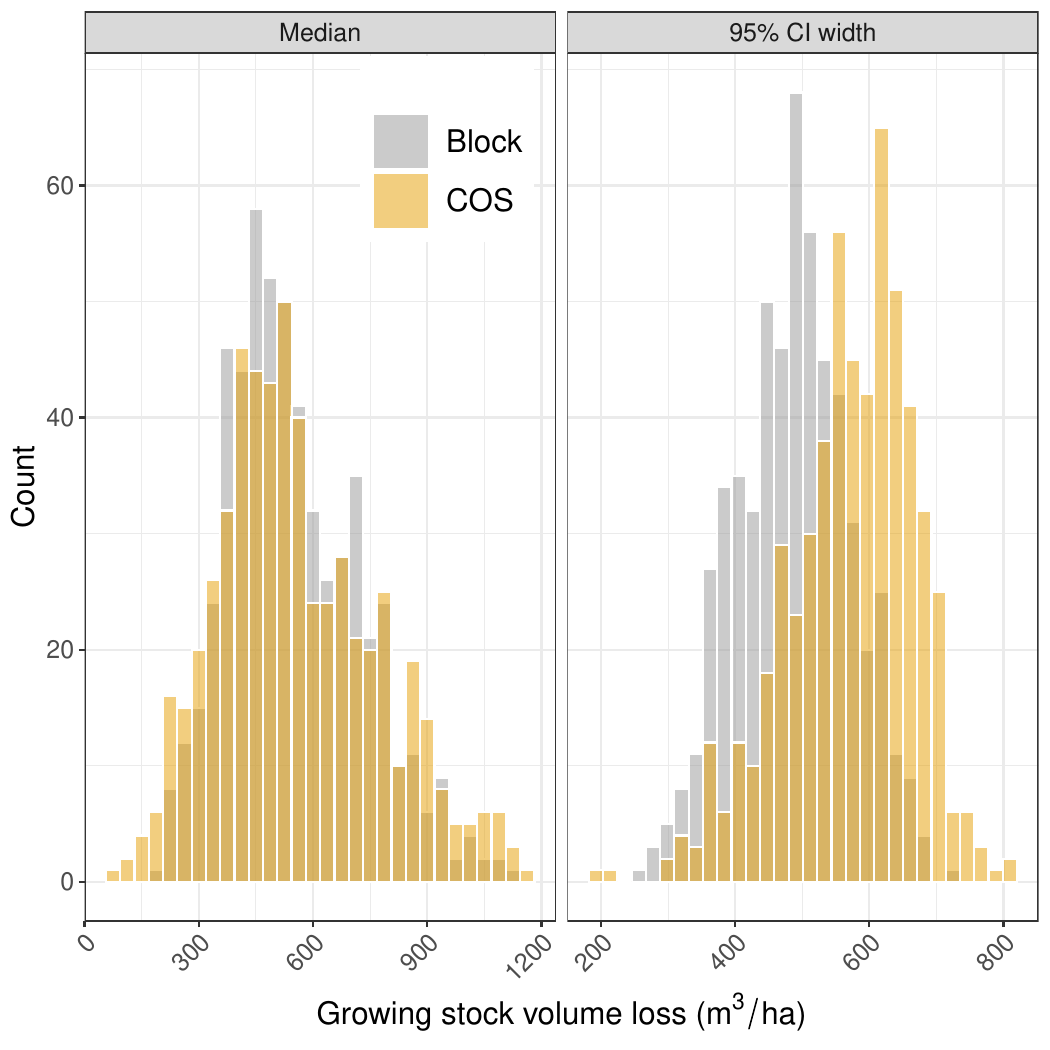}
	 \caption{Distribution of COS and Block growing stock volume loss posterior predictive distribution median and 95\% CI width for the 546 blowdowns. These are the distributions of point and interval estimates shown in Figure~\ref{fig:realCOSvsBlock}.} \label{fig:realHist}
\end{figure}

Figure~\ref{fig:realCOSvsBlock} compares COS and Block posterior predictive point and 95\% CI estimates for each of the 546 blowdowns. While point and interval estimates are similar, there are differences. First, the Block approach shows more shrinkage toward the mean than COS. This can be seen most readily at the axis extremes with COS points above the 1-to-1 line for lower values and below the line for larger values. Second, the COS 95\% CI widths are generally larger than the Block interval widths. Both these features are more apparent when looking at the distribution of posterior predictive medians and CI widths shown in
Figure~\ref{fig:realHist}. Here, the distribution of median point estimates for COS is wider than that of the Block estimates, and the 95\% CI widths for COS are substantially larger than those of the Block approach.

\begin{figure}[ht!]
	\centering
	\includegraphics[width=0.8\textwidth,trim={0.1cm 0cm 0.0cm 0cm},clip]{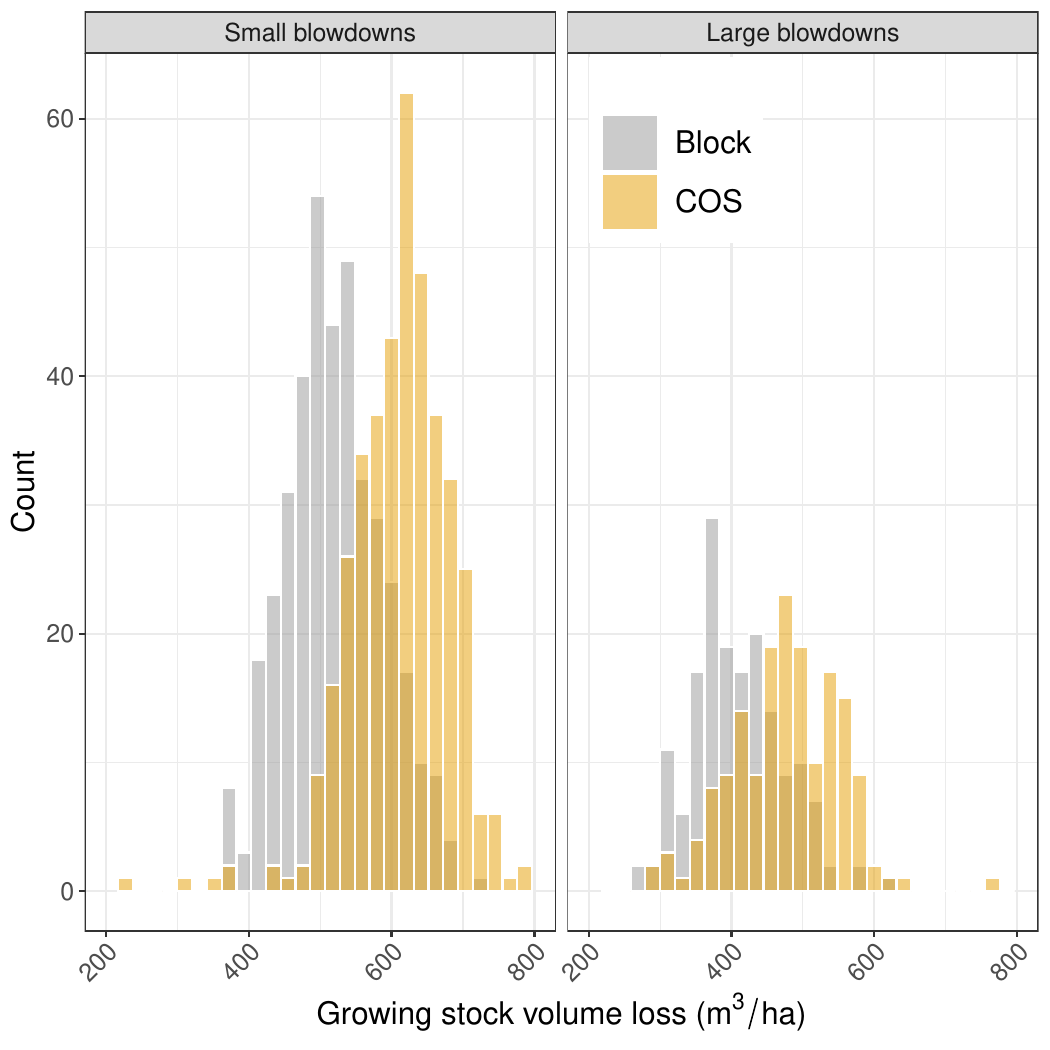}
	 \caption{Distribution of COS and Block growing stock volume loss posterior predictive distribution 95\% CI width for small area ($\le$ 0.25 ha) and large area ($>$ 0.25 ha) blowdowns.} \label{fig:realHistSmallLarge}
\end{figure}

Relative to the Block approach, COS informs prediction with more spatially detailed predictor variable information. Hence, one of our working hypotheses was that COS would deliver improved uncertainty quantitation, especially for smaller area blowdowns, and this improved uncertainty quantitation would likely come in the form of wider CI estimates (which we have seen in Figure~\ref{fig:realHist}). We also hypothesized that as the blowdown area increased, Block and COS would provide about the same uncertainty estimates. Indeed, as shown in Figure~\ref{fig:realHistSmallLarge}, COS generally delivers larger CI widths than Block for small blowdown areas (here we define ``small'' as approximately less than or equal to two times the observed inventory plot area); however, this figure also shows the trend of wider COS CIs extends to large blowdowns, but perhaps to a lesser degree. As expected, Figure~\ref{fig:realHistSmallLarge} also shows that as prediction unit area increases both methods deliver relatively smaller CI widths. 

\begin{figure}[!ht]
\begin{center}
	\includegraphics[width=0.8\textwidth,trim={0cm 3.5cm 0cm 3.5cm},clip]{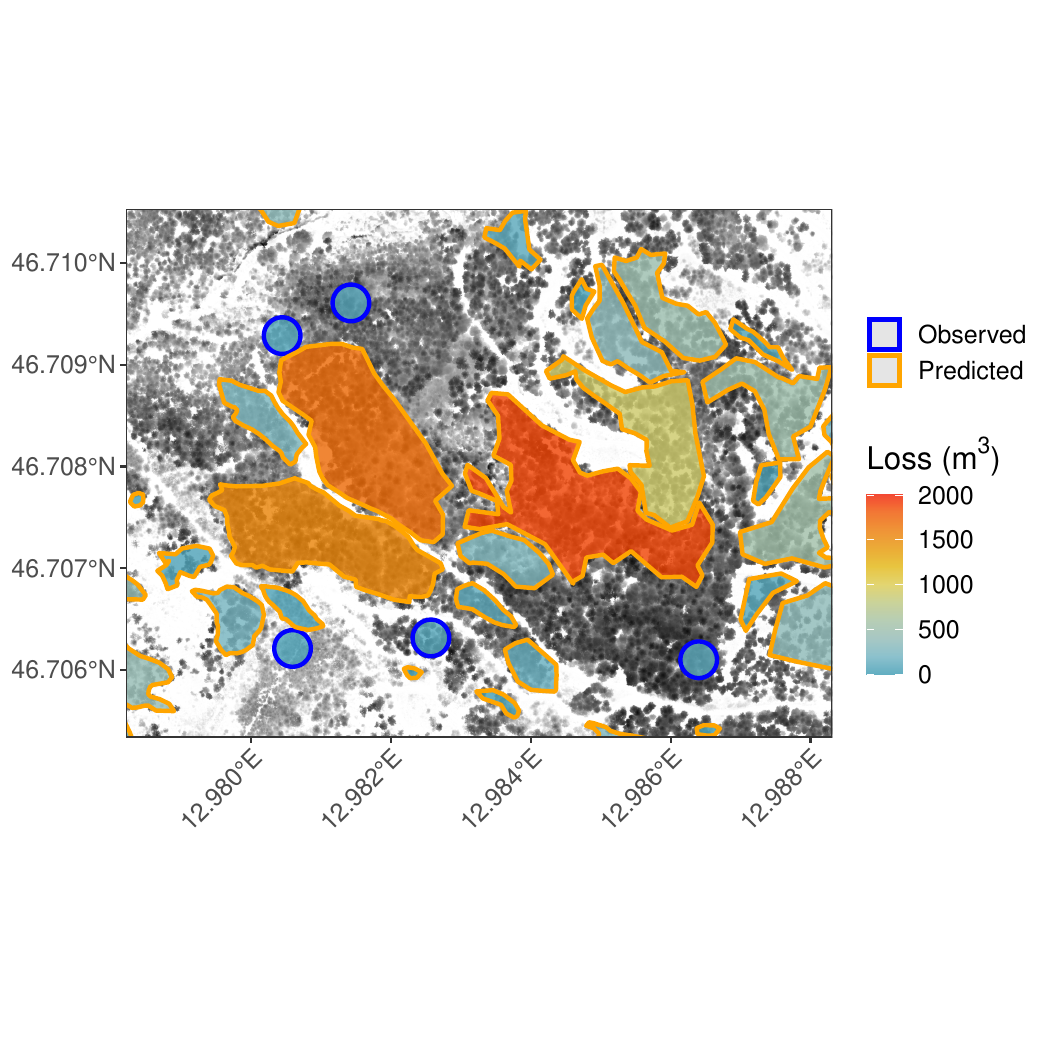}
	\caption{Observed inventory plots and COS growing stock volume loss posterior predictive distribution median for the blowdowns depicted in Laas sub-region Figure~\ref{plotAndPredictionUnitsLass}.} \label{plotAndPredictionsLass}
\end{center}
\end{figure}

As detailed in \citet{nothdurft2021estimating}, the Block approach provides separate---but spatially correlated via a joint predictive distribution---predictions for each prediction unit, i.e., large pixels, within a given blowdown. Posterior predictive distribution samples from each unit can be aggregated to arrive at a composite posterior predictive distribution for the given blowdown. In contrast, the COS approach delivers a single posterior predictive distribution for each blowdown as illustrated in Figure~\ref{plotAndPredictionsLass}. 

\begin{figure}[!ht]
	\begin{center}
		\subfigure[Median]{\includegraphics[width=0.4\textwidth,trim={2.25cm 0cm 2.25cm 0cm},clip]{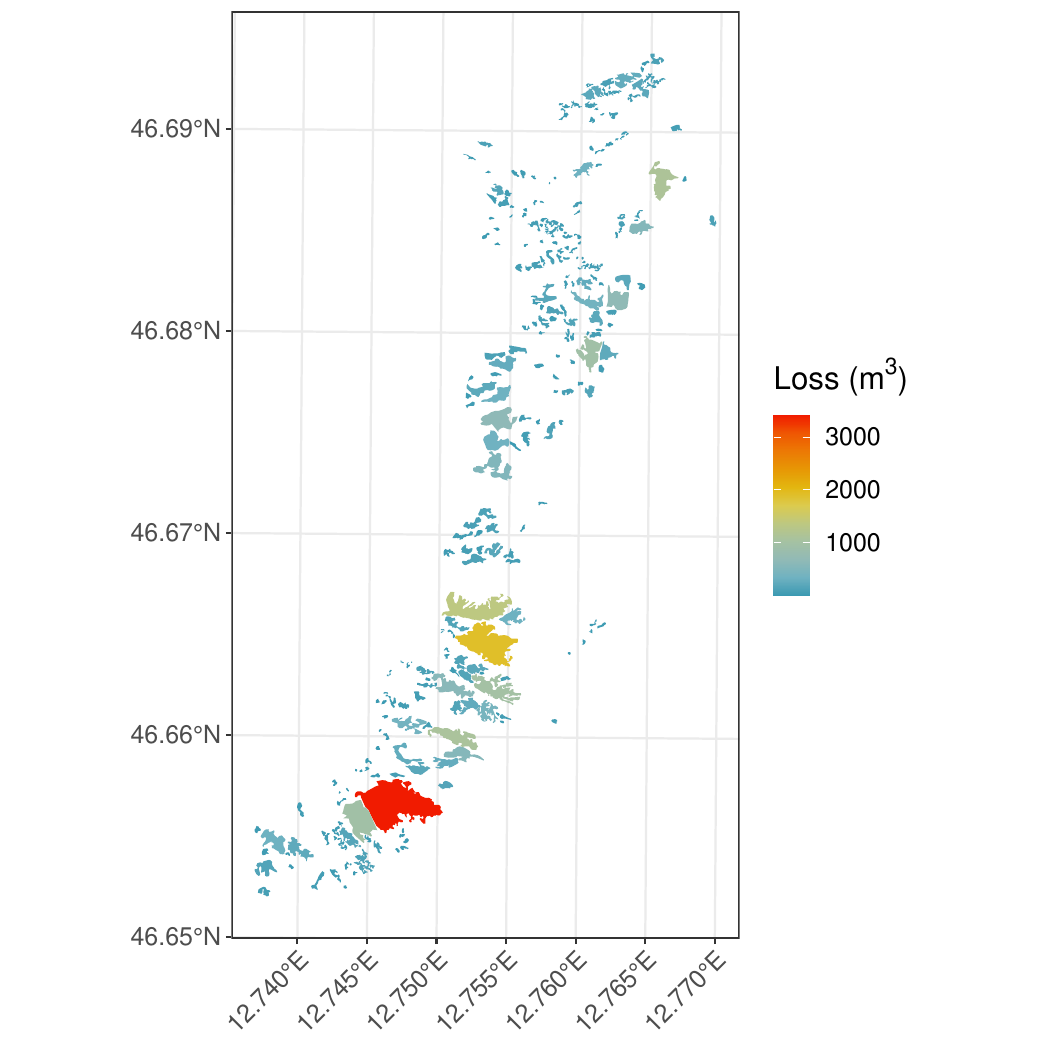}\label{fig:FrohnPredMedian}}
		\subfigure[95\% CI width]{\includegraphics[width=0.4\textwidth,trim={2.25cm 0cm 2.25cm 0cm},clip]{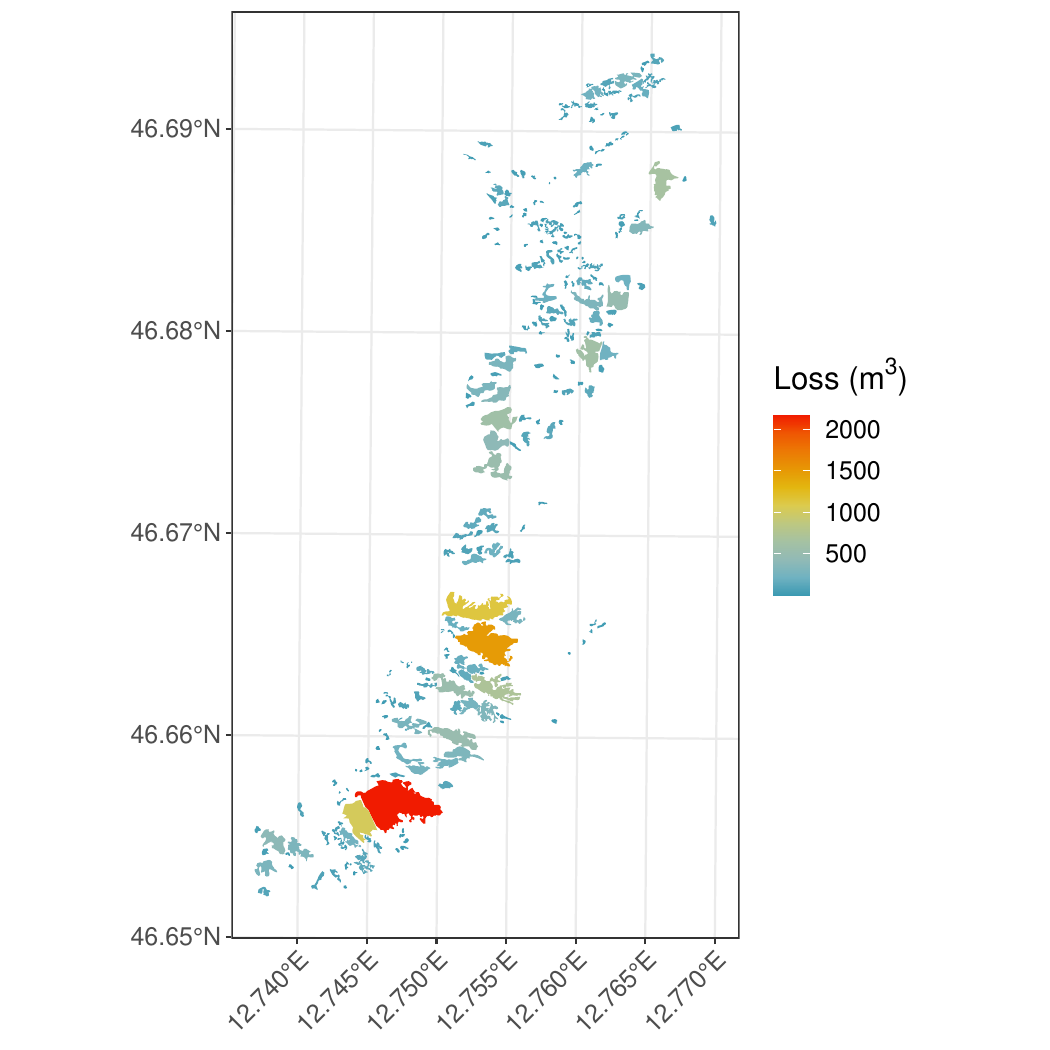}\label{fig:FrohnPredWidth}}
	\end{center}
	\caption{Total growing stock volume loss prediction summaries for blowdowns in Frohn sub-region. COS approach posterior predictive distribution median \subref{fig:FrohnPredMedian} and 95\% credible interval width \subref{fig:FrohnPredWidth}.} \label{fig:FrohnPred}
\end{figure}

COS prediction was performed for all blowdowns in each sub-region to produce maps useful for forest managers when assessing/quantifying storm blowdown damage, see, e.g., Figure~\ref{fig:FrohnPred}, and tabular summaries of sub-region and region totals given in Table~\ref{tab:predTotals}.

Consistent with blowdown-level prediction comparisons in Figures~\ref{fig:realCOSvsBlock} and \ref{fig:realHist}, Table~\ref{tab:predTotals} shows the two approaches produce comparable posterior predictive distribution median estimates for growing stock volume loss at sub-region and region levels. Table~\ref{tab:predTotals} also shows some change in uncertainty quantification when switching from Block to COS. With the exception of Lass, COS 95\% CI widths for the sub-regions and region totals are marginally larger than those for Block by 371.0, 213.6, 145.1, 67.2, and 486.5 cubic meters, for Frohn, Liesing, Mauthen, Ploecken, and region total, respectively. For Laas, Block 95\% CI was wider than COS by 358.4 cubic meters. 

\begin{table*}[ht!]
  \caption{Growing stock volume loss by sub-region and total region posterior predictive distribution median and 95\% credible interval.}\label{tab:predTotals}
  \begin{center}
  {\scriptsize
\begin{tabular}{lccc}
\toprule
  & Area (ha) & COS (m$^3$) & Block (m$^3$) \\
\midrule
Frohn & 62.83 & 31507.7 (27309.2, 35532.2) &31969.3 (28021.3, 35873.3)\\
Laas & 54.95 & 36869.7 (33142.7, 40310.4) &36146.4 (32530.9, 40057)\\
Liesing & 2.90 & 2422 (1870.5, 2871.2) & 2358.2 (1964, 2751.1)\\
Mauthen & 38.49 & 30884.8 (26982.9, 35586.5) & 30428.9 (26537, 34995.5)\\
Ploecken & 53.15 & 35122.7 (29387.2, 39348.2) &34613.1 (28848.2, 38742)\\
\midrule
Total & 212.32 & 136951.1 (125669.4, 145101.1) & 136486.3 (125203.2, 144148.4)\\
\bottomrule
\end{tabular}
}
\end{center}
\end{table*}

\section{Discussion and future work}\label{sec:discussion}
The near continuous advancement in sensor technology provides access to remotely sensed data of increased spatial resolution and extent. These high-resolution data are routinely coupled with other data streams of mixed resolution through various modeling techniques. Methods that accommodate these incompatible spatial data, such as the COS approach proposed here can yield improved inference in many settings and warrant additional development. 

In this article, we further explore models to assess forest damage from the 2018 Adrian storm in Carinthia, Austria. The study uses high-resolution LiDAR measurements and field measurements to quantify timber volume loss due to blowdown, which introduces the challenge of incompatible spatial data modeling. We address this challenge by proposing a new COS Bayesian method to leverage the rich information within high-resolution data. Our simulation study demonstrates the COS approach enhances prediction performance by effectively using the detailed information from high-resolution data. Compared to existing methods in forest damage assessment, our approach increases the resolution of the underlying model by approximately 140 times, leading to more expressive prediction uncertainty quantification at the individual blowdown-level. This significant enhancement in resolution underscores the potential of our proposed algorithm in transforming forest damage assessment and analysis.

Importantly, while the Block and COS approaches delivered substantially different uncertainty quantification at the blowdown-level, as seen in Figures~\ref{fig:realHist} and \ref{fig:realHistSmallLarge}, there was minimal inferential differences between the approaches when blowdown predictions were summarized at the sub-region and regional levels, i.e., as illustrated by small differences in total loss given in Table~\ref{tab:predTotals}. This suggests that if interest is in inference at the individual prediction units, e.g., blowdowns, then the COS approach should be favored, and either approach is a reasonable choice if one is interested in summaries over many prediction units, e.g., at the sub-region or regional levels. 

Our implementation of the COS approach utilizes covariance tapering to further reduce computational and storage costs. It is noteworthy that the proposed framework can be integrated with scalable Gaussian process algorithms that produce sparse or low-rank approximations of the covariance matrices, such as the Gaussian predictive process \citep{ban08} and the Multi-resolution Approximation \citep{katzfussmultires}. However, it is not compatible with scalable Gaussian process algorithms like the Nearest Neighbor Gaussian Process \citep{datta16} and other methods based on Vecchia's approximation \citep{ve88, guinness16, katzfuss2017general}, as these methods rely on a sparse precision matrix rather than a sparse or low-rank covariance matrix to achieve computational efficiency.

Looking ahead, the proposed algorithm offers potential for extension to spatio-temporal data modeling. This can be achieved by replacing the current model's spatial process with a spatio-temporal process. Additionally, exploring scalable modeling of the covariance function in the process model could yield further insights into algorithm efficiency, which we will leave for future studies.

\section*{Acknowledgments}

This work was supported by: S.B. NSF DMS-1916349, NSF DMS-2113778; A.N. Digi4+ and the Austrian Federal Ministry of Agriculture, Regions, and Tourism under project number 101470; A.O.F. NASA CMS grants Hayes (CMS 2020) and Cook (CMS 2018), NSF grant DMS-1916395, joint venture agreements with the USDA Forest Service Forest Inventory and Analysis, USDA Forest Service Region 9 Forest Health Protection Northern Research Station.  

\appendix
\section*{Appendix}\label{app: marginal_post}

Technical details for derivation of marginal posterior likelihoods \eqref{eq: marginal_posterior} \eqref{eq: marginal_post_flat_b}, and \eqref{eq: cond_posterior_B}. For \eqref{eq: marginal_posterior}, 
\begin{linenomath*}
\begin{align*}
&p(\sigma^2, \tau^2, \phi \given \by_B) = \int p(\bbeta, \bomega_B, \sigma^2, \tau^2, \phi \given \by_B ) d\bbeta d\bomega_B\\
&\propto \int \mbox{MVN}(\by_B \given \bomega_B + \bH_{BA} \bX \bbeta, \tau^2 \bD_h) \times \mbox{MVN}(\bomega_B \given 0, \sigma^2 \bC_B(\phi))\\
    & \quad \times \mbox{MVN}(\bbeta \given \bmu_\beta, \bV_\beta) d\bbeta d\bomega_B \times p(\tau^2, \sigma^2, \phi) \\
    &\propto \mbox{MVN}(\by_B \given \bH_{BA} \bX \bmu_\beta, \sigma^2 \bC_B(\phi) + \bH_{BA} \bX \bV_\beta \bX^\top \bH_{BA}^\top + \\
    &\quad \tau^2 \bD_h) \times p(\tau^2, \sigma^2, \phi)\;.
\end{align*}
\end{linenomath*}
For \eqref{eq: marginal_post_flat_b}, when assigning flat prior for $\bbeta$, i.e. $p(\bbeta) \propto 1$, we have
\begin{linenomath*}
\begin{align*}
&p(\sigma^2, \tau^2, \phi \given \by_B ) \propto \int p(\bbeta, \bomega_B, \sigma^2, \tau^2, \phi \given \by_B ) d\bbeta d\bomega_B\\
&\propto \int \mbox{MVN}(\by_B \given \bomega_B + \bH_{BA} \bX \bbeta, \tau^2 \bD_h) \times \\
&\quad \mbox{MVN}(\bomega_B \given 0, \sigma^2 \bC_B(\phi)) d\bbeta d\bomega_B \times p(\tau^2, \sigma^2, \phi) \\
    &\propto \int \mbox{MVN}(\by_B \given \bH_{BA} \bX \bbeta,\; \underbrace{\sigma^2 \bC_B(\phi) + \tau^2 \bD_h}_{\bV^\ast}) d\bbeta \times p(\tau^2, \sigma^2, \phi)\\
    &\propto \int \frac{1}{|\bV^\ast|^{1/2}}\exp[-\frac{1}{2}\{\bbeta^\top \underbrace{\bX^\top \bH_{BA}^\top \bV^{\ast-1}\bH_{BA}\bX}_{\bV_{\beta}^{\ast-1}} \bbeta - \\
    & \quad 2 \bbeta^\top \bX^\top \bH_{BA}^\top \bV^{\ast-1}\by_B + \by_B^\top \bV^{\ast-1} \by_B)\}]
    d\bbeta \times p(\tau^2, \sigma^2, \phi)\\
    &\propto \underbrace{\int \frac{1}{|\bV_{\beta}^\ast|^{1/2}}\exp[-\frac{1}{2}\{(\bbeta - \bmu_\beta^\ast)^\top \bV_{\beta}^{\ast-1}(\bbeta - \bmu_\beta^\ast) \}] d\bbeta}_{\text{const}}  \times \\
    &\quad \frac{|\bV_{\beta}^\ast|^{1/2}}{|\bV^\ast|^{1/2}} \exp[-\frac{1}{2} \{\by_B^\top \bV^{\ast-1} \by_B - \bmu_\beta^{\ast\top}\bV_{\beta}^{\ast-1}\bmu_\beta^{\ast} \}]\times p(\tau^2, \sigma^2, \phi)\\
    &\propto \frac{|\bV_{\beta}^\ast|^{1/2}}{|\bV^\ast|^{1/2}} \exp[-\frac{1}{2} \{\by_B^\top \bV^{\ast-1} \by_B - \bmu_\beta^{\ast\top}\bV_{\beta}^{\ast-1}\bmu_\beta^{\ast} \}]\times p(\tau^2, \sigma^2, \phi)\;,
\end{align*}
\end{linenomath*}
where 
\begin{linenomath*}
\begin{align*}
    \bmu_\beta^\ast &= \bV_\beta^{\ast} \bX^\top \bH_{BA}^\top \bV^{\ast-1}\by_B\;\\
    \bmu_\beta^{\ast\top}\bV_{\beta}^{\ast-1}\bmu_\beta^{\ast} &= \by_B^\top\bV^{\ast-1} \bH_{BA}\bX\bV_\beta^{\ast}\bX^\top \bH_{BA}^\top \bV^{\ast-1}\by_B\;.
\end{align*}
\end{linenomath*}
For \eqref{eq: cond_posterior_B}, $p(\bbeta, \bomega_B \given \sigma^2, \tau^2, \phi, \by_B )$
\begin{linenomath*}
\begin{align*}
     & \propto \exp( -\frac{1}{2\tau^2}\{(\by_B - \bomega_B -\bH_{BA}\bX\bbeta)^\top \bD_h^{-1}(\by_B - \bomega_B -\bH_{BA}\bX\bbeta)\} + \\
    &\frac{1}{2}(\bbeta - \bmu_\beta)^\top \bV_\beta^{-1}(\bbeta - \bmu_\beta) + \frac{1}{2\sigma^2}\bomega_B^\top \bC_B^{-1}(\phi)\bomega_B )\\
    &\propto \exp\left(-\frac{1}{2} 
    [\bbeta^\top, \bomega_B^\top] \bM^{-1}\begin{bmatrix} \bbeta \\ \bomega_B \end{bmatrix} + [\bbeta^\top, \bomega_B^\top]\mb\right) \sim \mbox{MVN}( \bM\mb, \bM)\;.
\end{align*}
\end{linenomath*}
where 
\begin{linenomath*}
\begin{align*}
 \bM^{-1} &= \left\{\begin{bmatrix}
    \bX^\top \bH_{BA}^\top \\ \bI_B \end{bmatrix} (\tau^2 \bD_h)^{-1} [\bH_{BA}\bX, \bI_B] + \begin{bmatrix} \bV_\beta^{-1} & 0 \\ 0 & \frac{1}{\sigma^2}\bC_B^{-1}(\phi) \end{bmatrix} \right\} \\  
\mb &= \left\{
    \begin{bmatrix} \bX^\top \bH_{BA}^\top\\ \bI_B \end{bmatrix}(\tau^2 \bD_h)^{-1}\by_B + \begin{bmatrix}  \bV_\beta^{-1}\bmu_\beta\\ 0 \end{bmatrix}   \right\}
\end{align*}
\end{linenomath*}

\bibliographystyle{elsarticle-harv} 
\bibliography{lubib}





\end{document}